\documentclass[%
 aip,
 rsi, long, numerical bibliography, (default)
 amsmath,amssymb,
 reprint,%
]{revtex4-1}

\usepackage{graphicx}% Include figure files
\usepackage{bm}% bold math
\usepackage{lineno} % do not change

\begin{document}
\preprint{AIP/123-QED}
%\preprint{}
\title{Low noise constant current source for bias dependent noise measurements } %Title of paper



\author{D. Talukdar}
\email{deep.talukdar@saha.ac.in}
%\thanks{}
\affiliation{Saha Institute of Nuclear Physics, 1/AF, Bidhannagar, Kolkata 700064}
\author{R. K. Chakraborty}
%\email{Second.Author@institution.edu.}
\affiliation{Bidhannagar College, EB - 2, Bidhannagar, Kolkata 700064}
\author{Suvendu Bose}
\affiliation{Saha Institute of Nuclear Physics, 1/AF, Bidhannagar, Kolkata 700064}
\author {K. K. Bardhan}
\affiliation{Saha Institute of Nuclear Physics, 1/AF, Bidhannagar, Kolkata 700064}

\date{\today}

\begin{abstract}
A low noise constant current source used for measuring the $1/f$ noise in disordered systems in ohmic as well as non-ohmic regime is described. The source can supply low noise constant current starting from as low as 1~$\mu$A to a few tens of mA with a high voltage compliance limit of around 20~Volts. The constant current source has several stages which can work in a standalone manner or together to supply the desired value of load current. The noise contributed by the current source is very low in the entire current range. The fabrication of a low noise voltage preamplifier modified for bias dependent noise measurements and based on the existing design available in the MAT04 data sheet  is also described.

\end{abstract}


\maketitle %\maketitle must follow title, authors, abstract and \pacs

\section{\label{sec:level1}INTRODUCTION }

				There has been an increasing interest in 1/$f$  noise phenomena in the last two decades. The reasons are mainly two-fold. Firstly, study of resistance fluctuations ($1/f$ noise) can be very informative for studying transport mechanisms in complex disordered systems like manganites\cite{Littlewood}, metal-insulator composites\cite{Wei,Aip}, CDW systems \cite{Subho} etc. Secondly, a device performs most satisfactorily if its noise level is within some specified limit thus making noise a sensitive measure of quality and reliability of the device\cite{Jevtic}.

In this article, we focus mainly on two kinds of situations that arise in noise measurements: a) One where resistance changes by a few orders of magnitude in certain materials undergoing phase transitions upon  varying a physical parameter like temperature\cite{uehera} or pressure. Examples include manganites of certain compositions where resistance changes abruptly with temperature; b) Another pertains to bias dependent noise measurement in disordered systems which are  driven into non-Ohmic ( i.e. non-linear ) regimes upon application of a bias of a few hundred mV or less \cite{Fuhrer,Gruner}. In this case, study of noise as a function of bias may be used to correlate the changes in the conduction mechanism or micro structure with the onset of nonlinearity. In addition to that, intriguing characteristics of noise in the non-linear regime is shown by some semiconductor devices like SCLC diode\cite{SCLC}, Schottky barrier diodes\cite{KLEINPENN} etc.

	   In conventional DC 4-probe setup for measuring resistance fluctuations (noise), a constant current, $I_{DC}=V_b /(R_b+r_s)$ from a battery ($V_b$) flows through a large ballast resistor ($R_b$) and the fluctuating sample ($r_s$). The average voltage across the sample, $\langle V \rangle $ = $ I_{DC} \langle   r_s \rangle$ is removed by AC coupling to a preamplifier so that voltage fluctuations $\delta V$ = $I_{DC} \delta R$  may be amplified and measured. Here, $\delta R$ stands for fluctuation in the resistance. The effect of contact fluctuations can be minimized by choosing $R_b/r_s$$\gg$1.  
Usually a wire wound or a metal film resistor 10 times the sample resistance with negligible level of flicker noise is used as the ballast. \cite{Scofield} 
		 % Clearly, bias dependent noise measurements require a battery operated low noise constant current source well isolated from noise sources such as computers and monitors and from 50Hz power line interferences.
  However, \emph{bias dependent} noise measurements require currents starting from a few $\mu$A to tens of mAs. Thus, one requires either an enormous number of batteries or huge values of noise free ballast resistors, both of which are unfeasible. Using a solid state current source connected to the mains also does not serve the purpose because of high level of noise introduced in the measurement chain from power line fluctuations. So a low noise constant current source satisfying the above requirements and having at least 15-20 V voltage compliance was needed. Few reported constant current sources in the literature \cite{Linzen,Ross} are for highly specific situations and not exactly applicable for our purpose. For instance, S. Linzen {\it et al.} \cite{Linzen} built a computer-controlled current source for quantum coherence experiments based on low noise operation amplifier OPA 627 which could however source a maximum current of only about 1.5mA and had voltage compliance limit of around 2.5~V. Another constant current supply for atomic CP violation experiments to drive a magnetic coil was constructed by Ross {\it et al.} \cite{Ross}. However, it could source only high value of currents and was optimized for low impedance samples. 
		
		In this article, we describe the design, operation and implementation of a simple low noise constant current source suitable for bias dependent noise measurements. By means of an accurate selection of active (JFET) and passive components (wirewound and metal film resistors) we have been able to design a versatile low noise constant current source with noise levels orders of magnitude lower than available commercial sources. The fabrication of a Bipolar Junction Transistor (BJT) coupled ultra low noise preamplifier based on the design given in Analog Devices MAT04 data sheet incorporating some modifications has been described in section II. Measurements using the constant current source and preamplifier is presented in section III.

\section{\label{sec:level2} SYSTEM DESIGN }

\subsection{Low noise constant current source}
	
%\subsubsection{Design objectives and component choice}
%		Characteristics of a good low noise current source are high output impedance, operability at low supply voltages, low temperature coefficient, adequate maximum operating voltage and requirement of minimum components. From this point of view, the Junction Field Effect Transistors (JFETs) become the natural choice as the basic component in fabricating the source. The main advantage of JFETs is that they are basically voltage-controlled devices which results in them having lower equivalent noise characteristics ($nV / \sqrt {Hz}$) than most bipolars\cite{Source}. Moreover, a low noise constant current source using a minimum number of components can only be constructed  with JFETs. 
%	

\begin{figure}
\begin{center}
\includegraphics[width=7cm]{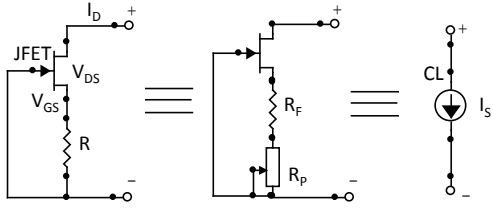}
\caption{Schematic of the basic module of the low noise constant current source.}
\label{fig.1}
\end{center}
\end{figure}

\begin{figure}
\begin{center}
\includegraphics[width=7cm]{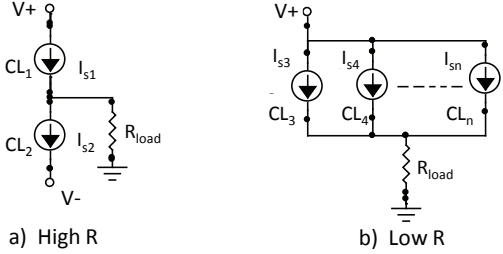}
\caption{Schematic of modules in different arrangements used for high and low current cases. a) High R module is for supplying low current while b) Low R module is for supplying high current.}
\label{fig.2}
\end{center}
\end{figure}

Figure 3 shows the circuit diagram of the current source. The design has two novel features, making it specially useful for bias dependent noise measurements. First, it has a modular construction where modules can be independently added or removed using switches to supply currents ranging over more than four decades.  Second, an arrangement has been made to keep the load floating above ground or referenced to ground for catering to different experimental situations arising in noise measurements.

\subsubsection{Design, circuit, and working }
	
In the `pinch off region' of a JFET, the drain current $I_D$ is almost independent of drain-source voltage $V_{\rm DS}$ but controlled by gate-source voltage $V_{\rm GS}$. Thus, JFET can be used as a current limiter (CL) when properly connected in a circuit\cite{Source} and is well documented in literature\cite{Siliconix}. The $I_D$ of the JFET in  Fig. 1 depends on the $V_{\rm GS}$ and the value of the resistor R. An increase in  $V_{\rm GS}$  reduces $I_D$ in the JFET, and any increase or decrease in $I_D$ causes similar changes in effective $V_{\rm GS}$  [$V_{\rm GS}$=$- R$ $I_D$]. This degenerative effect forces  $I_D$  to remain constant for a particular combination of R and $V_{\rm GS}$ . The use of R as self-bias for JFET enhances the JFET output impedance and the JFET behaves as CL. The resistor R can be series combination of a variable resistor $R_P$ and a fixed resistor $R_F$, and the total value ($R_P$ + $R_F$) can be adjusted for a particular requirement of $I_D$. Its value can be determined from the relation
\begin{equation}
R \, = \, \frac {V_{\rm GS}}{I_D} \, = \, \frac {V_{\rm GSOFF}}{I_D} \, \left (1-\sqrt {\frac {I_D}{ I_{\rm DSS}}}\right ).
\end{equation}

where $V_{\rm GSOFF}$ is the pinch off voltage and $I_{\rm DSS}$ is the pinch off current. The low frequency noise characteristics of the scheme described in Fig. 1 depends solely on the equivalent voltage noise source $S_{\rm v_{\rm FET}}$ of the JFET. The Power Spectral Density (PSD) of the current fluctuations in the load is given by\cite{Neri}:
\begin{equation}
	S_{i_{\rm FET}} \, = \, S_{ v_{\rm FET}} \left [\frac {g_{\rm m}}{1+g_{\rm m} R} \right]^2 . 
\end{equation}
where $S_{i_{FET}} $ is the PSD of fluctuation of drain current $i_{FET}$ through the JFET and $g_m$ is the transconductance.The voltage compliance of the scheme is limited by the maximum $V_{DS}$ of the JFET, i.e., the breakdown voltage.

\begin{figure*}
\begin{center}
\includegraphics[width=10.5cm]{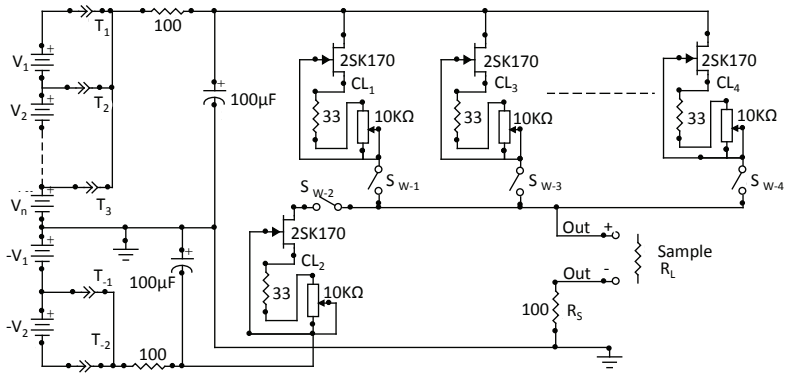}
\caption{Schematic of the low noise constant current source  consisting of four identical units as described in fig. 2. For details see text.}
\label{fig.3}
\end{center}
\end{figure*}

However, the scheme shown in Fig. 1 has some serious limitations while operating in low as well as high current regime. Noise measurement in low resistance samples requires high values of $I_D$ causing large power dissipation in the JFET  resulting in an increased temperature fluctuation of the active device. Consequently, the noise characteristics of the device degrade. As the amount of joule heating of the JFET depends on the product of $V_{DS}$ and $I_D$, this effect could be minimized by keeping $V_{DS}$ very low. The decrease in the value of $V_{DS}$ is restricted by the value of $V_{GSOFF}$ which ensures good current regulation. Again, for high resistance samples requiring low supply current, the FET adds significant voltage noise of its own~\cite{Datasheet}.
		
The problems associated with the basic design scheme of Fig. 1 has been solved using a novel circuit topology shown in Fig. 2. The circuit configuration shown in Fig 2a has been used for supplying low current to the sample. Here, two current limiter modules $CL_1$ and $CL_2$ connected in series, delivering currents $I_{S1}$ and $I_{S2}$ respectively, are used to deliver a low current ($I_{S1}$ - $I_{S2}$) through the sample $R_{load}$. The value of $I_{S1}$ is adjusted to be exactly $I_{load}$ amount more than that of $I_{S2}$ and the current through the $R_{load}$ is ($I_{S1}$ - $I_{S2}$). The $CL_1$ is connected to a positive voltage source (V+) and instead of connecting $CL_2$ to the ground, it is biased to a negative voltage source (V-). This ensures sufficient voltage drop across $CL_2$ which works as a constant current sink, even during a possible low voltage drop across the $R_{load}$. 		
		
Fig. 2b describes the circuit configuration for providing high current in few mA to tens of mA range. In this arrangement a high current through the sample $R_{load}$ can be passed using two or more current limiter modules $CL_{3-N}$, delivering currents $I_{S3}$, $I_{S4}$, - - - $I_{SN}$, connected in parallel. The supply voltage V+ can be derived from a battery of cells and the current limiters connected to the sample resistor work as a current source delivering a current of $I_{S3}$ + $I_{S4}$ +- - - + $I_{SN}$. 
				
Each of the CL's has a low noise n-type JFET [2SK170 manufactured by Toshiba] having typical $S_{v} = 0.95~nV / \sqrt {Hz} $  at f=1~KHz. A  variable resistor 10~K$\Omega$ and a 33~$\Omega$  combination has been used for the current control resistor. The variable resistances used are  Bourns-make wire wound resistances having negligible $1/f$ noise and 33~$\Omega$ resistors are excess noise free metal film resistances. The maximum current $I_S$  through each stage is selected to be $I_s\leq I_{dss(max)}/4$ to minimize the joule heating in the JFETs. Typically, the maximum current supplied through each stage is selected to be 5mA which makes the source capable of supplying multiples of 5~mA, depending on the number of CL's used.
%		When the right JFET with low noise and low $V_{GSOFF}$ is selected and is used in combination with low noise resistor for current control, each CL satisfies the requirements of a good low noise current limiter.
A combination of such CL's and filtered power from a battery of cells can be used for resistance fluctuations measurement.

		The complete current source scheme has been described in Fig. 3. It consists of different modules of CL's arranged in the manner described earlier. Each CL has been provided with one switch $S_{W-N}$. The number of CL's to be used can be selected with the right number N of $S_{W-N}$ to be ON. The supply voltage for the combination of CL's is derived from a battery of cells selected by the proper terminal $T_N$ and  low pass filters formed by 100~$\Omega$  resistor and 100~$\mu$F capacitor combination. The selection of supply voltage by $T_N$ depends on the correct $V_{DS}$ required for the combination delivering the selected $I_S$ keeping minimum power dissipated by the JFETs. The voltage drop across $R_S$ can be used to monitor the current delivered to the sample resistor. The placement of the current sensing resistor $R_S$ is critical as it can make the load referenced to ground or floating. In the source described above $R_S$ is placed in such a manner so as to make the load floating. This allows for noise measurement in differential mode which has the advantage of superior DC and AC common mode rejection resulting in increased signal to noise ratio (SNR) than single ended measurements which are susceptible to coupled noise and DC offsets. However, placement of $R_S$ and sample maybe interchanged for measurements with reference to ground, i.e., single ended one in place of differential.  

The module $CL_2$ can be connected to the combination of CL's using the  $S_{W1}$ and $S_{W2}$ for subtracting the current from $CL_1$ during low output current requirement. The right bias voltage $V_{-N}$ for $CL_2$ is selected by connecting through $T_{-N}$.

    The current source has been designed with minimum components without using any overall feedback loop. Such a loop would require more active and passive components and might well introduce additional $1/f$ noise. Besides, the drift of $I_S$ over the time of a spectral measurement, typically 5 minutes, is insignificant. The complete circuit, including the battery of cells, switches, etc., is enclosed in an electrically grounded sheet metal box to restrict influence of possible external noise. Thus, use of a number of CL's implemented as described can deliver the required range of current for the study of resistance fluctuations. 
	  
\subsection{Preamplifier}
	  
\subsubsection{Motivation}
 
	  It is customary to use a voltage preamplifier to amplify the voltage fluctuations of the sample before being fed to the spectrum analyzer.  Generally JFET or BJT input preamplifiers are preferred. The FET amplifiers generally work fine for samples having resistances above 10~K$\Omega$ owing to reasonable impedance matching with the high input impedance JFET of the input stage. However, impedance mismatch causes huge amount of noise to be added by the preamplifier leading to a large increase in the background noise\cite{Scofield} for samples having resistances below 10~K$\Omega$. BJT input preamplifiers are used to circumvent this problem of impedance mismatch for these low impedance samples. While excellent commercial FET preamplifiers such as SR560 by Stanford Research Systems or EG$\&$G PAR113 are available, it is not the case for BJT coupled preamplifiers. 
	  
     In bias dependent low frequency noise measurements the samples are often biased with large DC voltage. The preamplifier input requires AC-coupling down to very low frequencies ($f$ $\ll$ 1Hz) to reject the otherwise possible saturation of the high gain input stages of the preamplifier by large DC voltages\cite{Net81,Ciofi}. Generally, a coupling capacitor is used to pass through the AC fluctuations from the sample ($\delta$V) to the input of the JFET or BJT while DC bias is blocked. The lowest frequency ($f_c$) that is allowed through a combination of the input resistor R and the coupling capacitor C is given by:     
\begin{equation}
	    f_c=1/2\pi RC
\end{equation}
    In JFET input preamplifiers, presence of high input resistance suffices only a small value coupling capacitor to reduce possible attenuation even in mHz range. However, BJT input preamplifiers having low input resistance ($\approx$100~K$\Omega$) require a very high value coupling capacitor for working in the same frequency range. In addition to that, the high value coupling capacitor fails to block the DC voltage when the bias exceeds a few volts. Consequently, the frequency range of commercial BJT input preamplifiers starts from around 1~Hz only. The main cause of the problem is decrease in equivalent parallel resistance of the high value electrolytic capacitor on application of large voltage across it. This problem is solved by selecting a suitable non-electrolytic capacitor, as described below. 
     	  
   \subsubsection{Design Modification}
      
     A preamplifier based on the design given in the data sheet of MAT04 \cite{MAT04} has been fabricated to solve the problem discussed. It uses a precision dual matched transistor, the MAT02 and a feedback V-to-I converter, the MAT04. We used low noise operational amplifier OP27 instead of OP17 to act as an overall nulling amplifier to complete the feedback loop. The OP27 was used keeping in mind availability as well as the fact that huge bandwidth is not a requirement for our low frequency noise experiments.  We made a small modification at the input stage by using high value 120~$\mu$F, 250~V Solen \cite{Solen} metalized polyester capacitors instead of an electrolytic capacitor coupled with a 100~K$\Omega$ low tempco metal film resistor to form a RC coupling network. The Solen capacitors have low equivalent series resistance, low inductance, low dissipation factor along with high voltage rating of 250~V resulting in extremely low leakage currents thus providing superior low frequency coupling requirements. The preamplifier's frequency response starts from around as low as 100~mHz as a result of the modification. It also had a gain of 1000 and extremely low corner frequency of 3~Hz and voltage noise density of 2~$nV / \sqrt {Hz}$ at 10~Hz. The values are very close to those given in the data sheet of MAT 04.

\section{\label{sec:level3} RESULTS }

In order to characterize the source its current noise was determined for different values of load current. The current fluctuations at the output for a current $i_0$ passing through a load $R_L$ is given by
\begin{equation}
  \delta i = i_0 \frac {\delta R_{L}}{R_{L}} + \delta i_0^{(source)} 
\end{equation}
So the PSD of the current fluctuations $S_A$ is given by 
\begin{equation}
\label{eq:cur}
  S_A(f,i_0) = i_0^{2} \frac{S_{R_L}}{R_{L}^2} + S_{i_0}^{(source)} 
\end{equation}
where $S_{R_L}$ is the PSD for the resistance fluctuation of the load and  $S_{i_0}^{(source)}$ is the PSD for the current fluctuation of the source. According to Eq. (\ref{eq:cur}), the measured current noise approaches the current noise of the source as the load resistance $R_L$ is increased progressively, provided it has negligible flicker noise\cite{Art}. The noise $S_{i_0}^{(source)}$ of the JFET current source could be analyzed to have three possible noise sources: i) Shot noise \textit{ $2qi_0$}, ii) Johnson noise \textit{4kT/$R_S$ }, where \textit{$R_S$ }is the JFET source current setting resistor  and iii) JFET voltage noise. Fig. 4 shows typical values of $S_A(f,i_0)$ for a fixed load current of $i_o$=100$\mu$A using three different metal film resistors of values 150$\Omega$, 10K$\Omega$, 100K$\Omega$ as load. Still higher values of resistances could not be used as the output dc voltages reaches the compliance limit of the source for higher currents. 
The figure shows that measured current noise decreases with increasing load resistance in conformity with Eq. ($\ref{eq:cur}$). It also shows that we have already reached the noise floor for $R_L=100 K\Omega$. The low noise floor $S_{i_0}^{(source)}$$\approx$  $10^{-23} A^2/Hz$ of the source is lower than the shot noise values for the measured currents. The goal of fabricating a current source quieter than the shot noise was thus achieved which is the whole purpose of the JFET circuit. However it is to be noted that noise floor lower than $10^{-23} A^2/Hz$ could not be achieved as it is limited by the Johnson noise of the current setting resistor. Larger value of current setting resistor $R_S$ could be used to achieve low current as well as lower Johnson noise 4kT/$R_S$. However, it leads to an increase in the voltage noise of the JFET by orders of magnitude\cite{Datasheet} and also affects the stability of the load current. Thus for achieving lower currents, the circuit configuration described in fig 2a has been used where the difference of relatively higher currents ($\approx$ 1mA) flowing in opposite directions through two different JFET source modules were used to set the desired low current through the load. The higher currents were set by using  lower $R_S$ (typically $\approx$ 1 K$\Omega$) which resulted in an increase in the noise floor (johnson noise) $10^{-23} A^2/Hz$.
   
Fig 5 shows $S_A(f,i_0)$ for different $i_0$ for $R_L=150 \Omega$ and 10 K$\Omega$ respectively. A noise floor of $\approx$ $5 \times 10^{-23} A^2/Hz$ for low currents (up to 1mA) and  $\approx$ $3 \times 10^{-21} A^2/Hz$ for current over 10mA can be seen from fig. 5(b) and 5(a) respectively. Moreover, no unusual dependence of current noise with supplied current spanning over three decades has been found as is clear from the inset. The rise of the noise power at very low frequencies is due to the low frequency characteristics of the preamplifier\cite{Scofield} as can be confirmed by their noise figures (NF) for the particular source resistances and frequency.  

The $S_A(f,i_0)$ in ($\ref{eq:cur}$)  was calculated from voltage noise across the load $S_V(f,i_0)$. The experimental setup for measurement of $S_V(f,i_0)$ consisted of a suitably impedance matched low noise preamplifier (home made or PAR113) to amplify the voltage fluctuations $\delta$V produced in the sample. This was followed by a low pass anti aliasing filter (SR560) and finally to a dynamic spectrum analyzer (SR785) for evaluation of its spectral components. If $S_{vm}(f,i_0)$ be the measured PSD in spectrum analyzer then
\begin{equation}
 S_{vm}(f,i_{0})/G^2-S_{V}^{(B)}= S_V(f,i_0)
\end{equation}
where G is the gain of the preamplifier and $S_{V}^{(B)}$ is the background noise (amplifier + thermal noise).  Note that the 50Hz and its harmonics seen in fig. 5 and 6 are a result of the presence of the spectrum analyzer and the filter connected to the mains and is well known\cite{Linzen,Drung}. 

       A comparison of the rms noise power spectral density of a typical low noise constant current source like Keithley 6430 calculated from its data sheet\cite {Keithley} and present source is given in table I. It can be clearly seen from the table that except for the lowest current range, the noise performance of the described source is few orders of magnitude better than the commercial source Keithley 6430. 

\begin{table}[b]
\caption{\label{tab:table1}%
Comparison of current noise of a commercial source Keithley 6430 (noise calculated from data sheet) and the present source for different ranges of current. }
\begin{ruledtabular}
\begin{tabular}{ccc}

\textrm{Current}&
\textrm{Keithley 6430} &
\textrm{Present Source} \\
\textrm{range(A)} &
\textrm{$( A^2/Hz)$} &
\textrm{$( A^2/Hz)$} \\
\colrule
 10$\mu$A  & $2{\times}10^{-23}$ & $3{\times}10^{-23}$   \\ 
100$\mu$A  & $2{\times}10^{-21}$ & $5.1{\times}10^{-23}$   \\ 
1$~$mA  & $1.25{\times}10^{-18}$ & $6.2{\times}10^{-23}$     \\ 
10$~$mA  & $1.25{\times}10^{-16}$  & $3.5{\times}10^{-21}$     \\ 
\end{tabular}
\end{ruledtabular}
\end{table}

\begin{figure}
\begin{center}
\includegraphics[width=6.2cm]{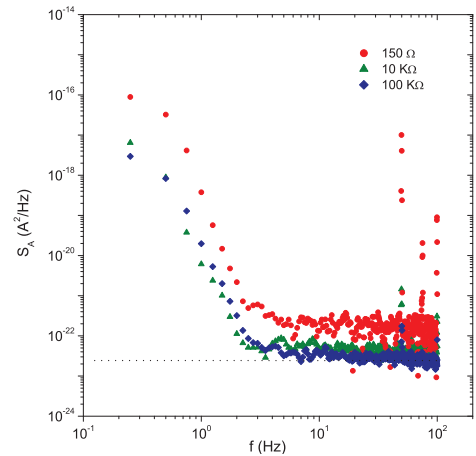}
\caption{Current noise $S_A$ vs. frequency for 3 different metal film resistors as load at one specific current of 100$\mu$A is shown. The trend of noise with load resistance follows eqn. $\ref{eq:cur}$.}
\label{fig.4}
\end{center}
\end{figure}

\begin{figure}
\begin{center}
\includegraphics[width=6.2cm]{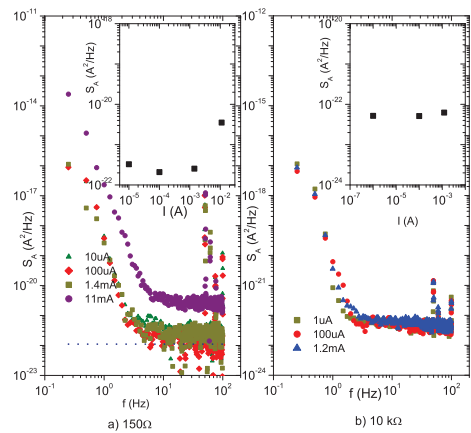}
\caption{Current noise $S_A$ vs. frequency for a 150~$\Omega$ and 10 K$\Omega$  metal film resistor for various bias currents are shown. The inset shows noise power (averaged over 9-18~Hz octave) with current. }
\label{fig.5}
\end{center}
\end{figure}

\begin{figure}
\begin{center}
\includegraphics[width=6.2cm]{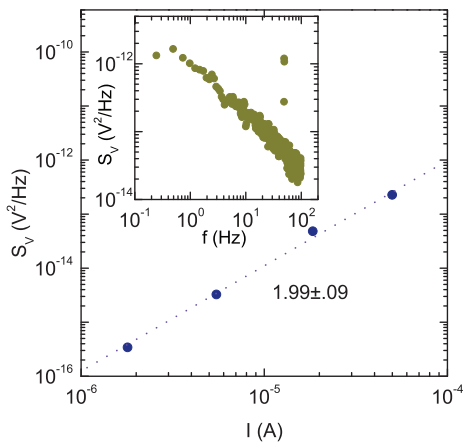}
\caption{PSD (averaged over 9-18~Hz) of a NSMO thin film with voltage at a temperature of 79~K. $V^2$ dependence of noise with bias is observed. A typical $1/f$ noise spectrum for the film is shown in the inset.}
\label{fig.6}
\end{center}
\end{figure}

In order to test the applicability of the source for noise measurements in homogeneous samples it is essential to verify the Hooge's law\cite {Hooge} which may be described as
\begin{equation}
S_v= 	\frac {\alpha V_{DC}^\beta} {N f^\gamma}
\end{equation}
where $V_{DC}$ is the average DC voltage drop across the sample, \textit{N} is the total number of charge carriers in a homogeneous sample, $\alpha$ is a dimensionless parameter depending upon the material and \textit{f} is the frequency. $\beta$ is the bias exponent generally equal to 2 for homogeneous samples and $\gamma$ is the frequency exponent having values in the range 0.8-1.3. Accordingly, low frequency noise measurements  were performed in a $Nd_{0.67}Sr_{0.33}MnO_3$ (NSMO) thin film\cite{Ramirez} of thickness 50nm deposited using Pulsed Laser Deposition (PLD) from a stoichiometric target onto $SrTiO_{3}$ (STO) single crystal substrates. After the deposition, thick gold layers were sputtered for making low resistance contacts to the sample. The film was then taken to a special SVT-400 Janis LHe cryostat and noise experiment performed at 78~K. The temperature was stabilized to better than $\pm10 mK$ during the measurement. Noise measurements were done using the current source and home made amplifier described earlier and the results shown in Fig. 6. Following Hooge's law a clear $V^2$ dependence of noise with current ranging over two decades in the ohmic regime was found.

\begin{figure}
\begin{center}
\includegraphics[width=6.2cm]{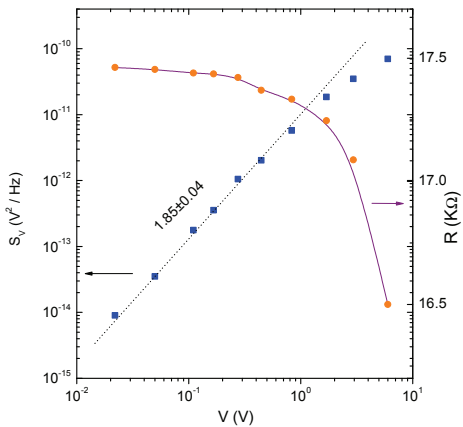}
\caption{Noise and Resistance with vs bias in a carbon wax composite. The bias exponent in the linear regime is 1.85. It can be clearly seen that the onset of non-linearity of resistance is accompanied with decrease of normalized noise. }
\end{center}
\end{figure}

		To show the usefulness of the source for bias dependent noise measurements extending up to non-linear regime, low frequency noise experiments were performed on carbon wax-composites. These materials are easily driven into non-linear regime when the voltage across them exceeds a few volts \cite{RKC} only. Using the constant current source and a JFET input PAR113 preamplifier, simultaneous measurement of noise and resistance as function of bias was made in carbon-wax composites and the results are presented in Fig. 7. In the figure it is clearly seen that the decrease of resistance with bias (onset of non-linearity) is picked up by the noise. The noise measurements have been performed over quite a large voltage range spanning over three decades. Bias exponent of noise in the linear range from 20~mV to 1.5~V is found to be around 1.85. This is in accordance with the behavior of bias exponent in disordered systems where  bias exponent of noise $\beta$ might differ\cite{Aip} from 2 even in the ohmic regime. Starting from around 1.5~V when the resistance starts becoming non-linear, the normalized noise magnitude also decreases. This leads to a further decrease of $\beta$ in the non-linear regime as is evident from the figure. The results of Fig. 7 establishes the relevance of bias dependent resistance noise measurement in  providing new insight into the conduction mechanisms of disordered systems extending up to the non linear regime. Using a battery and a ballast resistor as a current source for the above measurement would have required at least 70~V battery source at the highest value of applied voltage across the sample. Measurement of noise in a higher resistance sample would have required huge noise free ballast resistances not commonly available.

 \section{\label{sec:level4} CONCLUSIONS }
In this paper we have addressed the problem of the design of a very low noise constant current source using a very simple and modular circuit configuration capable of supplying currents ranging over a wide range with reasonably high voltage compliance. The flexible design of the source makes it suitable for noise measurements in situations where the resistance varies over a wide range in a single experiment or for bias dependent low frequency noise measurements. The noise performance of the proposed low noise constant current source is much better than those obtained by employing commercial instrumentation. Its greatest advantage lies in it's simple topology, very low cost and extremely low component count.

\begin{acknowledgments}
The authors would like to thank Dwijendra Das for his help in fabricating the prototypes and S. Mukhopadhyaya for providing the NSMO thin film. They would also like to thank the anonymous referees for very useful suggestions.
\end{acknowledgments}


 \end{document}